\documentclass[a4paper,twoside]{article}
\usepackage{amsmath,amssymb,amsfonts}
\pdfoutput=1
\usepackage{graphics}
\DeclareGraphicsExtensions{.pdf, .jpg}
\def\R{{\mathbb R}}

\date{ }
\title{Quantifying galactic clustering and departures from randomness of the inter-galactic
void probability function using information geometry\footnote{Presented at: Workshop on Statistics of Cosmological Data Sets NATO-ASI Isaac Newton Institute 8-13 August 1999. }}

\author{C.T.J. Dodson\\{\small\it School of Mathematics, University of Manchester,
  Manchester M13 9PL, UK}\\
  {\small\it ctdodson@manchester.ac.uk} }
  
\begin{document}
\maketitle
\begin{abstract}
A number of recent studies have estimated the inter-galactic void
probability function and investigated its departure from various
random models. We study a family of parametric statistical models
based on gamma distributions, which do give realistic descriptions
for other stochastic porous media. Gamma distributions contain as
a special case the exponential distributions, which correspond to
the `random' void size probability arising from Poisson processes.
The random case corresponds to the information-theoretic maximum
entropy or maximum uncertainty model. Lower entropy models
correspond on the one hand to more `clustered' structures or `more
dispersed' structures than expected at random. The space of
parameters is a surface with a natural Riemannian structure, the
Fisher information metric. This surface contains the Poisson
processes as an isometric embedding and provides the geometric
setting for quantifying departures from randomness and perhaps on
which may be written evolutionary dynamics for the void size
distribution. Estimates are obtained for the two parameters of the
void diameter distribution for an illustrative example of data
published by Fairall.
 MSC classes: 85A40; 60D05 

\end{abstract}

\section*{Introduction}
A number of studies over the past ten years
have estimated the inter-galactic
void probability function and investigated its departure
from randomness. The basic random model is that arising from a Poisson
process of mean density $n$ galaxies per unit volume in a large box.
Then, in a {\em given region of volume} $V,$ the probability of finding
exactly $m$ galaxies is
\begin{equation}
P_m =\frac{(nV)^m}{m!} e^{-nV} \label{poisson}
\end{equation}
So the probability that the given region is devoid of galaxies is
$P_0=e^{-nV}.$ Then it follows that the probability density function
for the continuous random variable $V$ in the Poisson case is
\begin{equation}
p_{random}(V) = n \, e^{-nV} \label{negexp}
\end{equation}
For comparison with observations, the approximation fails for
very large $V$  since a finite volume box is involved in any catalogue.

A hierachy of $N$-point correlation functions needed to represent
clustering of galaxies in a complete sense was devised by
White~\cite{white} and he provided explicit formulae, including
their continuous limit. In particular, he made a detailed study of
the probability that a sphere of radius $r$ is empty and showed
that formally it is symmetrically dependent on the whole hierarchy
of correlation functions. However, White concentrated his
applications on the case when the underlying galaxy distribution
was a Poisson process, the starting point for the present approach
which is concerned with geometrizing the parameter space of
departures from a Poisson process.

\section*{Geometry of gamma models for void volume statistics}
We choose a family of parametric statistical models that
includes~(\ref{negexp}) as a special case. There are of course
many such families, but we take one that has been successful in
modelling void size distributions in terrestrial stochastic porous
media~\cite{dodsonsampson} and has been used in the representation
of clustering of galaxies~\cite{ijtp}. The family of gamma
distributions has event space $\Omega=\R^+,$ parameters
$\mu,\beta\in \R^+$ and probability density functions given by
\begin{equation}
f(V;\mu,\beta)=\left(\frac{\beta}{\mu}\right)^\beta \,
\frac{V^{\beta-1}}{\Gamma(\beta)}\, e^{-V\beta/\mu}
\label{feq}
\end{equation}
Then $\bar{V}=\mu$ and $Var(V)=\mu^2/\beta$ and we see that $\mu$ controls
the mean of the distribution while the spread and shape is controlled
by, $1/\beta,$ the square of the coefficient
of variation.

\begin{figure}
\begin{picture}(300,220)(0,0)
\put(80,0){\resizebox{10cm}{!}{\includegraphics{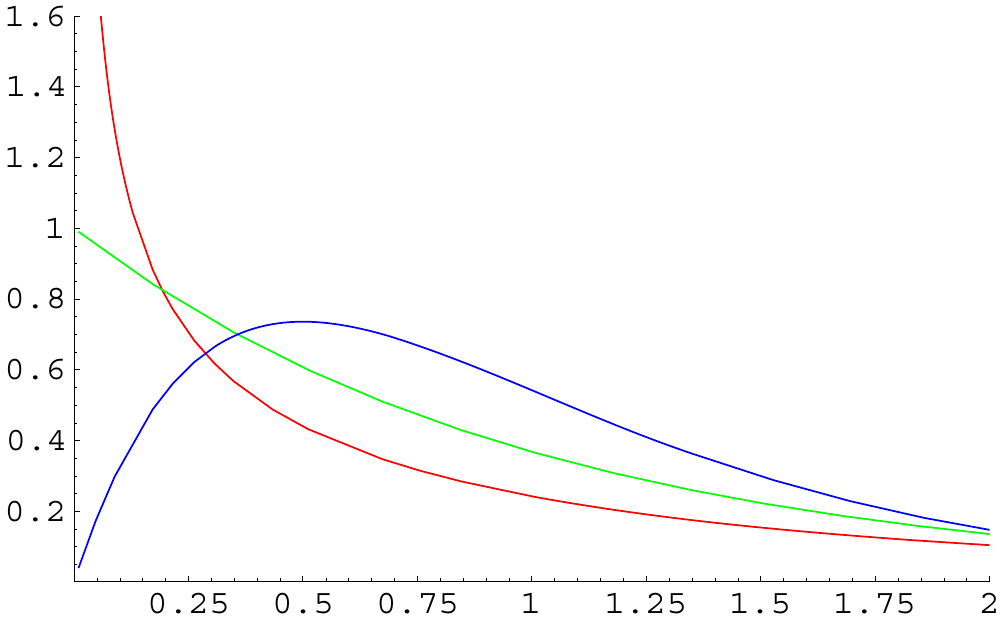}}}
\put(45,177){$f(V;1,\beta)$}
\put(125,160){$\beta=0.5$}
\put(180,70){$\beta=1$}
\put(310,50){$\beta=2$}
\put(300,-7){Void volume $V$}
\end{picture}
\caption{{\em Probability density functions, $f(V;\mu,\beta),$ for gamma
distributions of void volumes with unit
mean $\mu=1,$ and $\beta= 0.5, \ 1, \ 2.$ The case $\beta=1$ corresponds
to a `random' distribution from an underlying Poisson process.}}
\label{fgamma}
\end{figure}
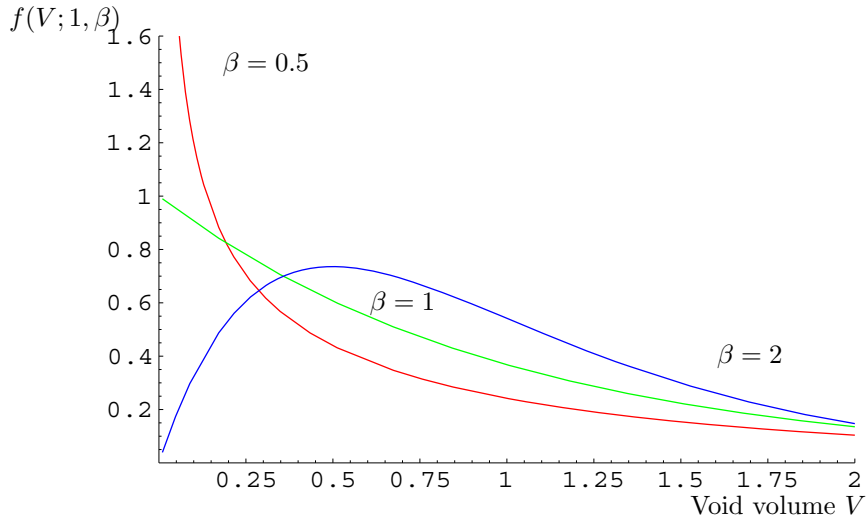

The special case $\beta=1$ corresponds to the situation when $V$
represents the random or Poisson process in~(\ref{negexp}) with
$\mu=1/n.$ Thus, the family of gamma distributions can model a
range of stochastic processes corresponding to non-independent
`clumped' events, for $\beta<1,$ and dispersed events, for
$\beta>1,$ as well as the random case
(cf.~\cite{dodson,dodsonsampson}). Thus, if we think of this range
of processes as corresponding to the possible distributions of
centroids of extended objects such as galaxies that are initially
distributed according to a Poisson process with $\beta=1,$ then
the three possibilities are:
\begin{description}
\item[Chaotic or random structure] with no interactions among
constituents, $\beta=1;$
\item[Clustered structure] arising from mutually attractive
interactions, $\beta<1;$
\item[Dispersed structure] arising from mutually repulsive
interactions, $\beta>1.$
\end{description}
Figure~\ref{fgamma} shows a family of gamma distributions, all of
unit mean, with $\beta= 0.5, \ 1, \ 2.$

Shannon's information theoretic `entropy' or `uncertainty' for such stochastic processes
(cf. eg. Jaynes~\cite{jaynes}) is given, up to a factor, by the
negative of the expectation of the logarithm of the probability density function~(\ref{feq}),
that is
\begin{eqnarray}
S_f(\mu,\beta) &=& -\int_0^\infty \log(f(V;\mu,\beta)\, f(V;\mu,\beta) \, dV \\
               &=& \beta+(1-\beta)\frac{\Gamma'(\beta)}{\Gamma(\beta)} +
\log\frac{\mu \, \Gamma(\beta)}{\beta} \label{fentropy}
\end{eqnarray}
In particular, at unit mean, the maximum entropy (or maximum uncertainty)
occurs at $\beta=1,$ which is the random case, and then $S_f(\mu,1)=1+\log\mu.$

The `maximum likelihood' estimates $\hat{\mu}, \hat{\beta}$ of $\mu,\beta$
can be expressed in terms of the mean and mean logarithm of a
set of independent observations $X=\{X_1,X_2,\ldots,X_n\}.$ These estimates are obtained
in terms of the properties of $X$ by maximizing the `log-likelihood' function
$$l_X(\mu,\beta)=\log lik_X(\mu,\beta)=\log\left(\prod_{i=1}^np(X_i;\mu,\beta)\right)$$
with the following result
\begin{eqnarray}
\hat{\mu}&=&\bar{X}=\frac{1}{n}\sum^n_{i=1}X_i\\ \log\hat{\beta}
-\psi(\hat{\beta})
   & = & \overline{\log X} - \log\bar{X}
\end{eqnarray}
where $\overline{\log X}=\frac{1}{n}\sum^n_{i=1}\log X_i$ and
$\psi(\beta)=\frac{\Gamma'(\beta)}{\Gamma(\beta)}$ is the digamma
function, the logarithmic derivative of the gamma function

The usual Riemannian information metric on the parameter
space $\cal{S}=\{(\mu,\beta)\in\R^+\times\R^+\}$
is given by
\begin{equation}
ds_{\cal{S}}^2=\frac{\beta}{\mu^2} \, d\mu^2 +
        \left(\psi'(\beta) -\frac{1}{\beta}\right)\, d\beta^2 \ \
        \ {\rm for} \ \mu,\beta\in\R^+ . \label{gammametric}
\end{equation}
For more details about the geometry
see~\cite{lauritzen,dodson,tg}. The 1-dimensional subspace
parametrized by $\beta=1$ corresponds to the available `random'
processes. A path through the parameter space $\cal{S}$ of gamma
models determines a curve
\begin{equation} c:[a,b]\rightarrow {\cal{S}}: t\mapsto
(c_1(t),c_2(t))\end{equation} with tangent vector
$\dot{c}(t)=(\dot{c}_1(t),\dot{c}_2(t))$ and norm $||\dot{c}||$
given via~(\ref{gammametric}) by
\begin{equation} ||\dot{c}(t)||^2=\frac{c_2(t)}{c_1(t)^2} \, \dot{c}_1(t)^2 +
        \left(\psi'(c_2(t))
        -\frac{1}{c_2(t)}\right)\,\dot{c}_2(t)^2.\end{equation}
The information length of the curve is \begin{equation}
L_c(a,b)=\int_a^b||\dot{c}(t)|| \, dt \end{equation} and the curve
corresponding to an underlying Poisson process has $c(t)=(t,1),$
so $t=\mu$ and $\beta=1=constant,$ and the information length is
$\log\frac{b}{a}.$

As we know from elementary geometry, arc length is often difficult
to evaluate analytically because it contains the square root of
the sum of squares of derivatives. Accordingly, we sometimes use
the `energy' of the curve instead of length for comparison between
nearby curves. Energy is given by integrating the {\em square} of
the norm of $\dot{c}$
\begin{equation}
E_c(a,b)=\int_a^b||\dot{c}(t)||^2 \, dt. \label{energy}
\end{equation}
so in the case of the curve $c(t)=(t,1),$ the energy is
$\frac{b-a}{ab}.$ It is easily shown that a curve of constant
$\mu$ has $c(t)=(constant,t)$ where $t=\beta$ and
$\dot{c}(t)=(0,1);$ this has energy
$\log\frac{a}{b}+\psi'(b)-\psi'(a).$

Locally, minimal paths joining nearby pairs of points in $\cal{S}$
are given by the autoparallel curves or geodesics~\cite{tg}
defined by~(\ref{gammametric}). Some typical sprays of geodesics
emanating from various points are provided in~\cite{ijtp}. The
Gaussian curvature of the surface $\cal{S}$~\cite{tg} actually
controls all of the geometry of geodesics and it is given by
\begin{equation}
K(\mu,\beta)= {\frac{\psi'(\beta) +\beta \psi''(\beta)  }
{4(\beta\psi'(\beta)-1 }}\ \ \
        \ {\rm for} \ \mu,\beta\in\R^+
\end{equation}
\begin{eqnarray}
K_{\cal{S}}(\mu,\beta)&\rightarrow& -\frac{1}{4} \  {\rm as} \
\beta\rightarrow 0\\ K_{\cal{S}}(\mu,\beta)&\rightarrow& -\frac{1}{2} \
{\rm as} \ \beta\rightarrow \infty
\end{eqnarray}

\section*{Void diameter statistics}
For a general account of large-scale structures in the universe,
see Fairall~\cite{fairall}.
Kauffmann and Fairall~\cite{kauffmannfairall} developed a catalogue
search algorithm for nearly spherical regions devoid of bright galaxies
and obtained a spectrum for diameters of significant voids. This had
a peak in the range 8-11 $h^{-1}Mpc,$ a long tail stretching at least to
64 $h^{-1}Mpc,$ and is compatible with the recent extrapolation models
of Baccigalupi et al~\cite{bacc} which yield an upper bound on void diameters
of about 100 $h^{-1}Mpc.$ We shall return to the data of Kauffmann and Fairall
later in this section.

Simulations of Sahni et al.~\cite{sahni} found strong correlation between
void sizes and primordial gravitational potential at void centres; void
topologies tended to simplify with time. Ghigna et al.~\cite{ghignabr} found in
their simulations that void statistics are sensitive to the passage from
CDM to CHDM models. This suggested that the void distribution is sensitive
to the type of dark matter but not to the transfer function between types.
CHDM simulations gave a void probability in excess of observations,
CDM simulations being somewhat better. Vogeley et al.~\cite{vogeley}
compared void statistics with CDM simulations of a range of cosmological
models; good agreement was achieved for samples of very bright galaxies
($M<-19.2$) but for samples containing fainter galaxies the predicted
voids were reported to be `too empty'. Ghigna et al.~\cite{ghignabb} compared
observational data with Gaussian-initiated $N$-body simulations
in a $50 \, h^{-1} Mpc$ box and found that at the $2-8 \, h^{-1} Mpc$
scales the void probability for CHDM was significantly larger than observed.
Ghigna et al.~\cite{ghignabt} compared simulated galaxy samples with the
Perseus-Pisces redshift survey. The void probability function did
discriminate between DM and CDM models, the former giving
particularly good agreement with the survey.

Little and Weinberg~\cite{little} used similar $N$-body simulations,
and found that the void probability was insensitive to the shape of the
initial power spectrum. Watson and Rowan-Robinson~\cite{watson}
found that standard CDM predictors do yield reasonably good void
probability function estimates whereas Voronoi foam models performed
less well.

Lachi\'{e}ze-Rey and daCosta were of the opinion that the available
samples of galaxies were insufficiently representative, because of greater
apparent frequency of larger voids in the southern hemisphere.
Bernardeau~\cite{bernardeau} started from a Gaussian field and derived
an expression for the void probability function, obtaining
$$\frac{\log P_0}{nV} \approx -(nV \, \sigma^2)^{-3/7} \ \ {\rm for \
large} \ nV \, \sigma^2,$$
with $\sigma^2$ the variance of the number of galaxies in volume $V.$
This distribution has an extended large tail because it is asymptotically
more like the exponential of $-(nV)^\frac{4}{7}$ than the Poisson case
which decays like the exponential of $-nV.$
Cappi et al.~\cite{cappi} examined the dependence of the void probability
function on scale for a range of galaxy cluster samples, finding a general
scaling to occur up to void diameters of about $100 \, h^{-1} Mpc.$
Kerscher et al.~\cite{kerscher} used the void probability function to
obtain spatial statistics of clusters on scales $10-60 \, h^{-1} Mpc,$
obtaining satisfactory agreement in a model with a cosmological constant
and in a model with breaking of scale invariance of perturbations.

\begin{figure}
\begin{picture}(300,220)(0,0)
\put(80,0){\resizebox{10cm}{!}{\includegraphics{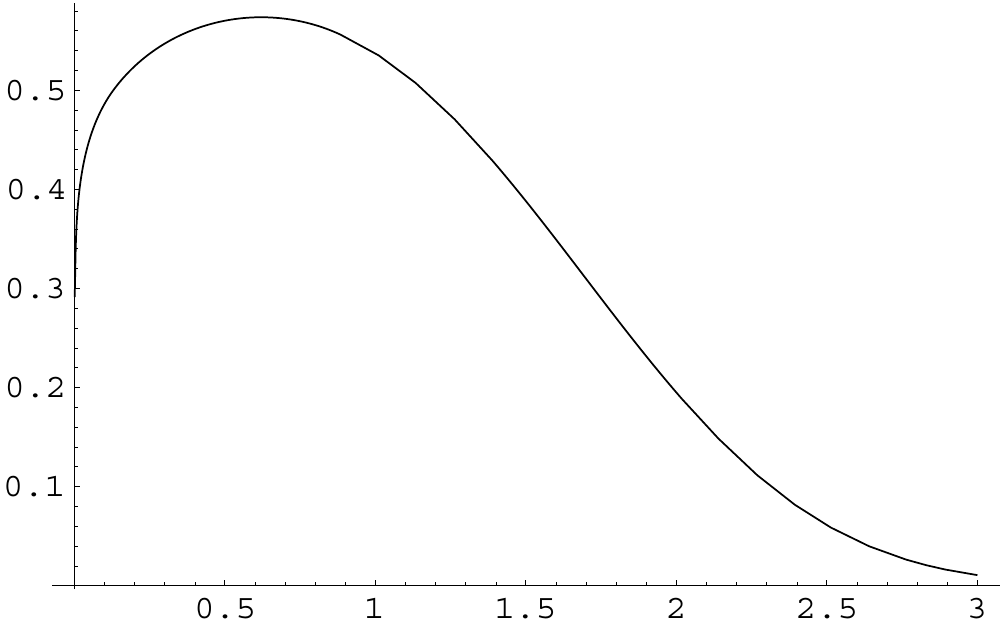}}}
\put(43,160){$h(D;\mu,\beta)$}
\put(300,-7){Void diameter $D$}
\end{picture}
\caption{{\em Probability density function Equation~\protect(\ref{Dpdf}), for
distributions of void diameters with unit
mean, $\mu=1.244$ and $\beta=0.370.$ These parameter values are the best
fit for the data of Kauffmann and
Fairall~\protect\cite{kauffmannfairall}, also used in
Figure~\protect\ref{kfzcat}.}}
\label{hpdf}
\end{figure}

For our model we consider the diameter $D$ of a spherical void with
volume $V=\frac{\pi}{6} D^3$ having distribution~(\ref{feq}).
Something close to the random variable $D$ has direct representation in some
theoretical models, for example as polyhedral diameters in Voronoi
tesselations~\cite{weygaert,weygaerticke,coles}.

\begin{figure}
\begin{picture}(300,220)(0,0)
\put(50,0){\resizebox{12cm}{!}{\includegraphics{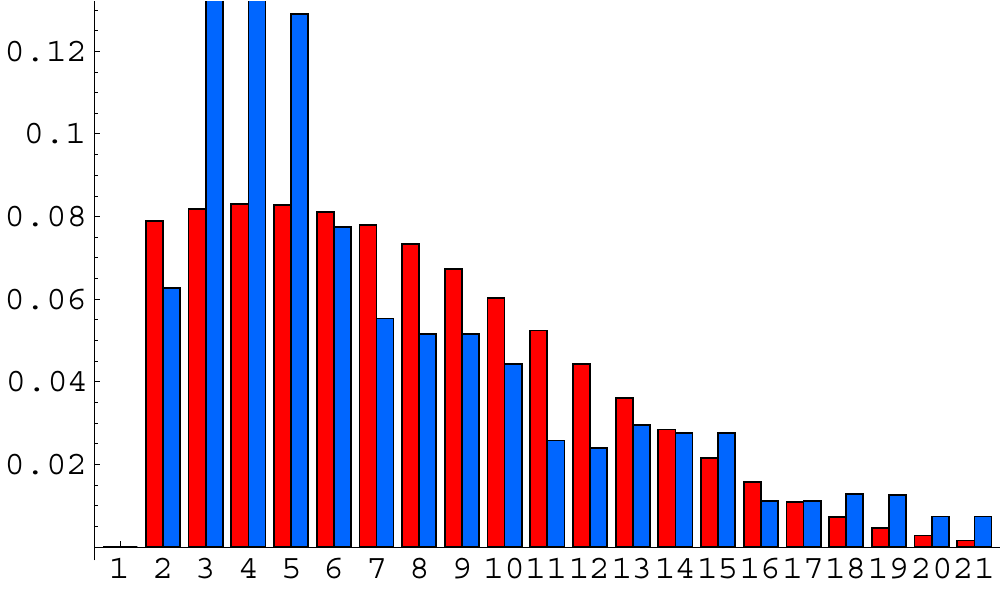}}}
\end{picture}
\caption{{\em Histograms of fractions of space occupied by different classes
of void diameters, from Kauffmann and
Fairall~\protect\cite{kauffmannfairall} (right hand columns)
and predicted from the void diameter
distribution Equation~\protect(\ref{Dpdf}) (left hand columns)
fitted to the
same coefficient of variation and mean; the parameters found were
$\mu=1.244$ and $\beta=0.370.$  The class centres are in units of $200 \,km/s$
with a mean close to 7 units.}}
\label{kfzcat}
\end{figure}

The probability density function for $D$ is given by
\begin{equation}
h(D;\mu,\beta)= f(\frac{\pi}{6} D^3;\mu,\beta)\, \frac{dV}{dD} = \frac{1}{\Gamma(\beta)}\,
\left(\frac{  \beta }{ \mu }\right)^\beta \,
\left(\frac{ \pi D^3}{6}\right)^{\beta-1} \, e^\frac{ -\pi\beta D^3}{6\mu}
\label{Dpdf}
\end{equation}
Then the mean $\bar{D},$ variance $Var(D)$ and coefficient
of variation $CV(D)$ of $D$ are given, respectively, by
\begin{eqnarray}
\bar{D}&=& \frac{
 \Gamma(\beta+\frac{1}{3})   }
{  \left(\frac{\pi\beta}{6\mu}\right)^\frac{1}{3} \Gamma(\beta)  }
\label{meanD} \\
Var(D)&=&  \frac{
    \Gamma(\beta) \, \Gamma(\beta+\frac{2}{3})     -\Gamma(\beta+\frac{1}{3})^2   }
{  \left(\frac{\pi\beta}{6\mu}\right)^\frac{2}{3} \Gamma(\beta)^2  }\label{varD} \\
CV(D)&=&\sqrt{
\frac{\Gamma(\beta) \, \Gamma(\beta+\frac{2}{3})}
{\Gamma(\beta+\frac{1}{3})^2}-1 } \label{cvD}
\end{eqnarray}

The fact that the coefficient of variation~(\ref{cvD}) depends
only on $\beta$ gives a rapid fitting of data to~(\ref{Dpdf}).
Numerical fitting to~(\ref{cvD}) gives $\beta;$ this substituted
in~(\ref{meanD}) yields an estimate of $\mu$ to fit a given
observational mean. By way of illustration, this has been done in
Figure~\ref{kfzcat} for the ZCAT/SRC data from Kauffmann and
Fairall~\protect\cite{kauffmannfairall,fairall} Figure 8a (cf
also~\cite{fairall} Figure 6.5), both set to unit mean
diameter---the true mean for that catalogue was about $7\, h^{-1}
Mpc.$ The fitted values were $\mu=1.244$ and $\beta=0.370.$ This
fit is not particularly good if the reported peaks are not an
artifact but, qualitatively, we observe that the fitted value
$\beta=0.370$ is apparently significantly less than $1,$ which
would correspond to the random model. We conclude tentatively
that, for the ZCAT/SRC data subjected to the Kauffmann and Fairall
search algorithm, the new model suggests clumping rather than
dispersion in the underlying stochastic process. A {\em
Mathematica} NoteBook for performing the fitting procedure is
available from the author, who would welcome more sets of data.

Suppose that the parameter space $\cal{S}$ of gamma-based models is
a meaningful representation of the evolutionary process
and that some coordinates such as $(\mu=1.244,$ $\beta=0.370)$
in $\cal{S}$ represent current data. Then geodesics in $\cal{S}$
through this point represent some kind of extremal path.
Moreover, we may consider a vector field $U$ on $\cal{S}$ such that
$(\mu=1.244,$ $\beta=0.370)$ is the present endpoint of an integral curve of
$U,$ the initial point of this curve being presumably in an epoch when
less clustering ($\beta>0.370$), a random state ($\beta=1$) or even dispersion
($\beta>1$) was present at higher density.
It would be interesting to investigate the various candidate
cosmologies for their appropriate vector fields, via the statistics
of matter and voids they predict. The present gamma-related parametric statistical
models provide the means to convert the catalogue statistics into
the coordinate parameters and a background geometrization of the statistics
on which dynamical processes may be formulated.

\section*{Acknowledgement} The author wishes to thank A.P. Fairall, F. Labini
and M.B. Ribeiro for helpful comments during the preparation of
this article.

\end{document}